%% file: manuscript.tex
\documentclass{ifacconf}
\usepackage{xspace}
\usepackage{soul}
\usepackage{mathtools}
\usepackage{makecell}
\usepackage{nicefrac}
\usepackage{float}
\usepackage{wrapfig}
\usepackage{amssymb}
\usepackage{color}
\usepackage{amsmath}
\usepackage{siunitx}
\usepackage{cuted}

\newcommand{\ie}{\unskip, i.\,e.,\xspace}
\newcommand{\eg}{\unskip, e.\,g.,\xspace}

\newcommand{\sut}{\text{s.\,t.\,}}

\newcommand{\N}{\ensuremath{\mathbb{N}}}

\newcommand{\X}{\ensuremath{\mathbb{X}}}

\newcommand{\U}{\ensuremath{\mathbb{U}}}

\makeatletter
\newcommand{\subalign}[1]{%
	\vcenter{%
		\Let@ \restore@math@cr \default@tag
		\baselineskip\fontdimen10 \scriptfont\tw@
		\advance\baselineskip\fontdimen12 \scriptfont\tw@
		\lineskip\thr@@\fontdimen8 \scriptfont\thr@@
		\lineskiplimit\lineskip
		\ialign{\hfil$\m@th\scriptstyle##$&$\m@th\scriptstyle{}##$\crcr
			#1\crcr
		}%
	}
}

\usepackage{tcolorbox}

\usepackage{todonotes} % murp added for notes support
\usepackage[inline, shortlabels]{enumitem}
\usepackage{graphicx}      % include this line if your document contains figures
\usepackage{natbib}        % required for bibliography

\begin{document}
\begin{frontmatter}

\title{Model Predictive Control of a Food Production Unit: A Case Study for Lettuce Production
	%Model Predictive Control of Closed Biomass Production Container: A Case Study for Lettuce Production%\thanksref{footnoteinfo}
} 
%\title{Cost adaptation over a prediction horizon: the feasibility and stability aspect %\thanksref{footnoteinfo}

\thanks[footnoteinfo]{This project has received funding from the European Social Fund (ESF)}

\author[First]{Murali Padmanabha} 
\author[First]{Lukas Beckenbach} 
\author[First]{Stefan Streif}

\address{Technische Universtit\"{a}t Chemnitz,
	Automatic Control and Systems Dynamic Lab, 09107 Chemnitz, Germany}
\address[First]{(e-mail:\{murali.padmanabha, lukas.beckenbach, stefan.streif\}@etit.tu-chemnitz.de)}
\begin{abstract}                % Abstract of not more than 250 words.
Plant factories with artificial light are widely researched for food production in a controlled environment. 
For such control tasks, models of the energy and resource exchange in the production unit as well as those of the plant's growth process may be used. 
To achieve minimal operation cost, optimal control strategies can be applied to the system, taking into account the availability of resources by control reference specification. 
A particular advantage of model predictive control (MPC) is the incorporation of constraints that comply with actuator limitations and general plant growth conditions.
In this work, a model of a production unit is derived including a description of the relation between the actuators' electrical signals and the input values to the model.
Furthermore, a preliminary model based state tracking control is evaluated for production unit containing Lettuce. 
It could be observed that the controller is capable to track the reference while satisfying the constraint under changing weather conditions and resource availability.
%For such systems, a dynamical model of the energy and mass exchange in the chamber is derived and extended to include dynamics of photosynthesis and respiration of the plant, allowing model based control approaches for plant growth. This model includes a description of the relation between the actuators' electrical signals and the input values to the model.
%In this work, a preliminary model predictive controller (MPC) based state tacking for climate control is evaluated on the closed biomass productions container with Lettuce. 

\end{abstract}


\end{frontmatter}
%===============================================================================
\graphicspath{{./gfx/}}
\section{Introduction}

Conventional controlled environment agriculture based on greenhouses are transforming into highly sophisticated plant factories for continuous food production. These plant factories, also referred to as indoor-vertical-farms and sometimes as urban-farms, utilize infrastructures (warehouses, shipping containers, etc.) with artificial light and precisely controlled climate for production of biomass (e.g. plants, fish, algae) \citep{KOZAI2013}. Regardless of the efficient use of resources (water, land, etc.), sustainability of these farms are eminently criticised due to the high energy consumption for artificial lighting and climate and inefficient byproduct reuse \citep{ALCHALABI201574,GRAAMANS201831}.

A recent study on economically feasible vertical farms by \cite{CONRAD2017} has shown potential for sustainable operation by incorporating multiple production units (plant-unit, fish-unit, etc.) and interconnecting them for byproduct reuse. % \citep{CONRAD2017,Mario2019}.
Such interconnection imparts additional complexity to the efficient control of production units which are inherently nonlinear due to the underlying biological processes that depend on several states (temperature, humidity, CO$_2$, water etc.) and also the conglomerate of actuators that influences them. Therefore, for efficient operation of such farms, it is necessary to consider resource (electricity, CO$_2$ etc.) change dynamics due to external disturbances together with state and input constraints. 

Works of \cite{Henten1994,VANSTRATEN2000221} propose the use of optimal control approaches to manage greenhouse climate (temperature and humidity) for plant production demonstrating economic benefits using mathematical models and weather forecast data.
In particular, model predictive control (MPC) is a widely applied optimal control approach that utilizes a process model, possibly including disturbance specifications, and considers input and state constraints as well as economic factors to track desired references (states and/or inputs) \citep{Kim2002,Ferreau2007,Gu2006}.

A particular focus of the current investigation is to set up a suitable predictive control scheme for a plant biomass production unit under the following considerations:
\begin{enumerate*} [1)]
\item resource availability from other production units (CO$_2$, H$_2$O, etc.),
\item changes in weather,
\item actuator limitations and operation costs,
\item relevant state constraints for plant survival, and
\item dynamics not included in the plant model but necessary for plant growth (day-night pattern)
%\item external influences necessary for plant growth (day-night pattern)
\end {enumerate*}.

In this work, a detailed mathematical model describing various dynamics of a production unit is presented along with the hardware limitations and effects of the disturbance on the system's states. % such as actuator capacity and input signal range.
A dynamical model of the plant growth presented in literature is considered for the subject produced in the production unit.
References for states representing the optimal growth conditions, disturbances (resource availability and weather) and input references for light intensity are specified.
%The references for inputs (LEDs) are specified explicitly to compensate for the dynamics not modelled.
%Using these fixed reference, the performance of tracking MPC is analyzed under the influence of disturbance.
Given these references, control operation of a tracking MPC is evaluated and the influence of the disturbance is investigated. 
Although disturbances may pose a distinct difficulty in predictive controlling, in general, a robustness analysis for the controller is bypassed by employing specific disturbance functions over the prediction horizon. 
%While such, in general unknown, disturbances pose a distinct difficulty in controlling, the required robustness analysis for the controller is bypassed by employing specific disturbance functions over the prediction horizon. 
These functions are available through short-term weather forecasts and render the process model time-variant.

The following Section \ref{sec:modeling} consolidates the model equations of the system under study.
Then, in Section \ref{sec:control}, the online optimization of the predictive controller is introduced for climate tracking.  
Section \ref{sec:result} evaluates the performance of the predictive controller while discussing certain issues related with the reachability of the reference. 
A conclusion and outlook is given in Section \ref{sec:conclusion}. 
 
\input{modelling}
\input{control}
\input{results_discussion}

\section{conclusion}\label{sec:conclusion}

This work reviews the application of a nonlinear predictive controller on a growth chamber for climate tracking control. 
%Specifically, a dynamical model for the container-plant-environment interaction was considered, with state trajectories given for temperature, humidity and CO$_2$ concentration. 
Specifically, a disturbance affected prediction model of the container-plant-environment interaction was considered, in which particular trajectories have been substituted for the disturbance. 
Additionally, references for temperature, humidity and CO$_2$ concentration and light intensity have been specified, representing best growth conditions. 

The controller is applied on the nonlinear system, relaxing the binary input constraint to ease the computation. 
It has been observed that the (heuristic) optimal plant growth environment could be tracked within the limitation of the actuators, while using a short horizon in combination with large sensor and actuator sampling. 
%Due to these limitations, however, certain errors in tracking could not be circumvented under the considered outside climate. 
As for the nonlinearity of the model, further investigations could consider utilizing adaptive control methods in finding efficient control actions that require less computational resources.
%The end goal being the development of a MPC based controller running on a resource constrained embedded PC for tracking given state (climate and resource) trajectories such that the container operates energy efficiently.
%In a future work, a MPC based controller running on a resource constrained embedded PC for tracking given state (climate and resource) trajectories shall be developed.%, such that the container operates energy efficiently.
In a future work, the MPC based controller developed in this work shall be implemented to run on a resource constrained embedded PC.% for tracking given state trajectories.
\bibliography{bib/MPC,bib/ADP_RL,bib/ClassicOptControl,bib/NonlinearControl,bib/ConstructiveMath,bib/MPCappl_extra,bib/PlantModel}
%\appendix
%\section{A summary of Latin grammar}    % Each appendix must have a short title.
%\section{Some Latin vocabulary}              % Sections and subsections are supported  
%                                                                         % in the appendices.
\end{document}

%% file: modelling.tex
\section{Production Unit and Plant Model} \label{sec:modeling}

\begin{figure*}[h]
	\begin{center}
		%\includegraphics[width=13cm]{gfx/Plant_process_growth_chamber_ext}    % The printed column width is 8.4 cm.
		%		\resizebox{\linewidth}{!}{\input{./gfx/Plant_process_growth_chamber_ext.pdf_tex}}
		\def\svgwidth{\linewidth}
		\fontsize{7.5}{8}\selectfont
		\resizebox{0.87\linewidth}{!}{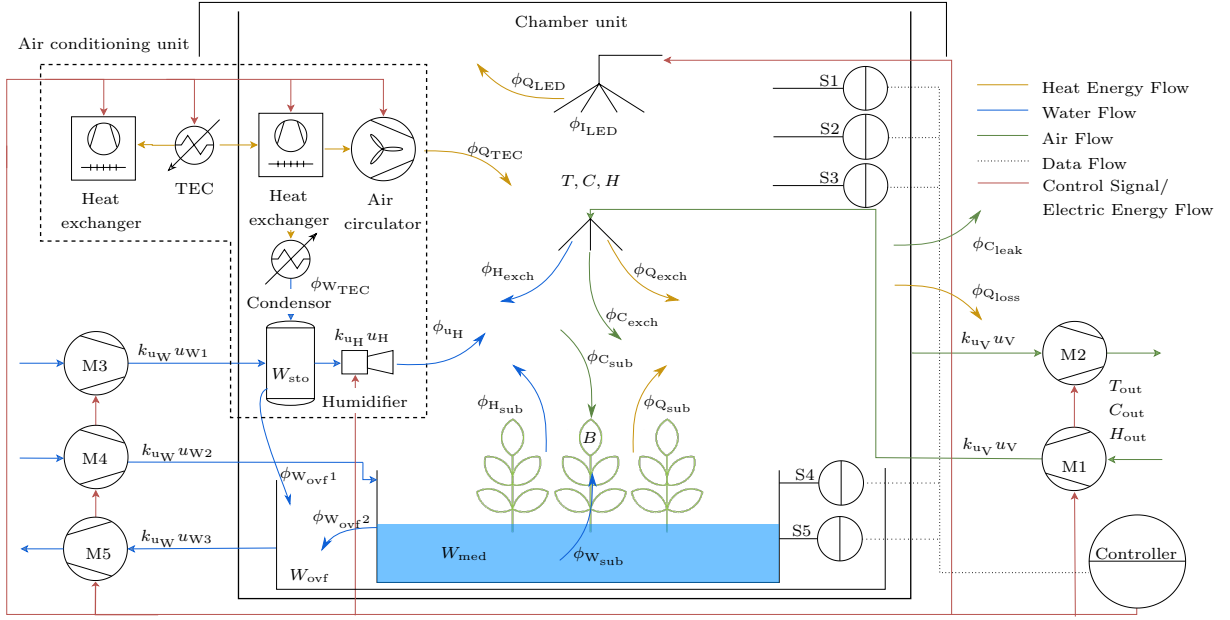}
		\caption{System components and resource flux: Production unit with sensors (S1-S5), air pumps (M1-M2), water pumps (M3-M5), air conditioning unit (thermoelectric cooler based for heating, cooling, condensation), humidifier, LED lighting, the corresponding system states, and some of the resource fluxes.} 
		\label{fig:growth_chamber}
	\end{center}
\end{figure*}

A prototype version of a production unit was developed in a scale comparable to a standard commercial growth chamber to serve as a test-bench \citep[see][for more details]{Padmanabha2019}. The designed controlled environment is integrated with various sensors and actuators to facilitate the regulation of climate and resource exchanges as depicted in Fig.~\ref{fig:growth_chamber}. 
Although several actuators are in place, controllability in the state space is restricted by limitations of the actuators and the influence of the external environment.
One such limitation is the cooling capacity of the thermoelectric cooler (TEC).

\subsection{Chamber model}\label{chamber_model}
Mathematical model of the mentioned growth chamber is derived as mass and energy balance equations based on work proposed for green house dynamics modeling \citep{STRATEN2011}.
%This growth chamber can be viewed as a system with inputs that signals the actuators to facilitate energy and mass flux, affecting various state variables of the system. Interaction of the subject growing inside these growth chambers with its environment can be regarded as internal disturbances and the change in outside climatic conditions are considered as external disturbances.
Details of the derived model with its various mass and energy flux components are presented in this section.
\begin{table}[h]
	\caption{List of important symbols}
	\centering	
	\begin{tabular}{lll}
		\hline
		Symbol                     & Description                                & Unit                           \\ \hline
		$T$                        & temperature of air inside chamber          & [\si{\celsius}]                \\
		$C$                        & CO$_2$ concentration of air inside chamber & [\si{\kilogram\meter}$^{-3}$]  \\
		$H$                        & absolute humidity of the air inside        & [\si{\kilogram\meter}$^{-3}$]  \\
		$W_\mathrm{sto}$           & total water in the storage tank            & [\si{\kilogram}]               \\
		$W_\mathrm{med}$           & total water in the growing medium          & [\si{\kilogram}]               \\
		$W_\mathrm{ovf}$           & total water overflowing                    & [\si{\kilogram}]               \\
		$B$                        & biomass/dry matter content of the crop     & [\si{\kilogram\meter}$^{-2}$]  \\
		$u_\mathrm{T}$             & TEC input                                  & [-]                            \\
		$u_\mathrm{V}$             & ventilator input                           & [-]                            \\
		$u_\mathrm{H}$             & humidifier input                           & [-]                            \\
		$u_\mathrm{W1}$            & storage tank pump input                    & [-]                            \\
		$u_\mathrm{W2}$            & growing medium pump input                  & [-]                            \\
		$u_\mathrm{W3}$            & overflow pump input                        & [-]                            \\
		$u_{\mathrm{I}i}$          & light input of $i^\mathrm{th}$ LED channel & [-]                            \\
		$T_\mathrm{out}$           & temperature of outside air                 & [\si{\celsius}]                \\
		$C_\mathrm{out}$           & CO$_2$ concentration of external source    & [\si{\kilogram\meter}$^{-3}$]  \\
		$H_\mathrm{out}$           & absolute humidity of external source       & [\si{\kilogram\meter}$^{-3}$]  \\ \hline
	\end{tabular}
\end{table}

\subsubsection{Heat Flux}
The temperature inside the chamber, represented by the state variable $T$, is affected by various heat fluxes.
Heat can be supplied to and removed from the chamber through the heater-cooler system.
A simplified equation presented by \cite{VIAN2002407} is used for modeling the heat flux term contributed by the TEC module
\begin{equation}\label{eq:Q_in}
%\phi_{\mathrm{Q_{TEC}}} =\frac{  k_\mathrm{\alpha,q} k_\mathrm{V,m} u_\mathrm{T} }{k_\mathrm{R,q}} \ T + \frac{(u_\mathrm{T} k_\mathrm{V,m})^2}{2 k_\mathrm{R,q}} + k_\mathrm{TEC}(T_\mathrm{out}-T),
\phi_{\mathrm{Q_{TEC}}} =\frac{  k_\mathrm{\alpha} k_\mathrm{V} u_\mathrm{T} }{k_\mathrm{R,q}} \ T + \frac{(u_\mathrm{T} k_\mathrm{V})^2}{2 k_\mathrm{R,q}} + k_\mathrm{q}(T_\mathrm{out}-T),
\end{equation}
where $k_\mathrm{\alpha}$, $k_\mathrm{R,q}$, $k_\mathrm{q}$, and $k_\mathrm{V}$ are the Seebeck coefficient, series resistance, and thermal conductivity and maximum operation voltage of the TEC module respectively.
The LED panel inside the chamber generates both heat and radiant flux and can be modeled as
\begin{equation}\label{eq:Q_led}
\phi_{\mathrm{Q_{LED}}}= \sum_{i=1}^{4}k_{\mathrm{Q,m}i}u_{\mathrm{I}i},\quad I = \sum_{i=1}^{4}\eta_{\mathrm{LU}i}k_{\mathrm{I,m}i} u_{\mathrm{I}i},
\end{equation}
where $i$ represents the narrow and wide-band wavelengths (LED channels) supported by the light panel, $k_{\mathrm{Q,m}i}$ and $k_{\mathrm{I,m}i}$ are the maximum heat and radiant light dissipated by the respective LED channel.

The heat flux components due to ventilation and leakage/conduction are expressed respectively as
\begin{equation}
%\begin{split}\label{eq:Q_comp}
%\phi_{\mathrm{Q_{exch}}}&=k_\mathrm{c,air}k_\mathrm{\rho,air}(T_\mathrm{out}-T)k_\mathrm{u_{v}}u_\mathrm{v}, \\
%\phi_{\mathrm{Q_{loss}}}&=k_\mathrm{a,chm}k_\mathrm{U,chm}(T_\mathrm{out}-T),
%\end{split}
\phi_{\mathrm{Q_{ex}}}=k_\mathrm{c}k_\mathrm{\rho}(T_\mathrm{out}-T)k_\mathrm{u_{v}}u_\mathrm{v}, \ \
\phi_{\mathrm{Q_{lo}}}=k_\mathrm{A}k_\mathrm{U}(T_\mathrm{out}-T),
\end{equation}
where $k_\mathrm{\rho}$ and $k_\mathrm{c}$ is the density and specific heat capacity of air respectively, $k_\mathrm{A}$ is the chamber's surface area, $k_\mathrm{U}$ is the coefficient of heat transfer through the walls, and $k_\mathrm{u_{v}}$ is the flow rate of the ventilator pump.

Finally, the rate of change of temperature in the chamber can be modeled as energy balance equation:
\begin{equation}\label{eq:dT/dt}
%k_\mathrm{C,chm}\frac{\mathrm{d}T}{\mathrm{d}t}=\phi_{\mathrm{Q_{exch}}}+\phi_{\mathrm{Q_{loss}}}+\phi_{\mathrm{Q_{TEC}}}+\phi_{\mathrm{Q_{LED}}}+ \phi_{\mathrm{Q_{sub}}},
k_\mathrm{C,chm}\dot{T}=\phi_{\mathrm{Q_{ex}}}+\phi_{\mathrm{Q_{lo}}}+\phi_{\mathrm{Q_{TEC}}}+\phi_{\mathrm{Q_{LED}}}+ \phi_{\mathrm{Q_{sub}}},
\end{equation}
where $k_\mathrm{C,cham}$ is the total heat capacity of the chamber and $\phi_{\mathrm{Q_{sub}}}$ represents the heat absorbed due to evapotranspiration.

\subsubsection{CO$_2$ and O$_2$ Flux}
The concentration of CO$_2$ and O$_2$ inside the chamber can be modeled as mass flux equations.
The influx due to ventilation and outflux due to leakage is given respectively as 
\begin{equation}\label{eq:phi_cexch}
\phi_\mathrm{C_{exch}} = (C_\mathrm{out}-C)k_\mathrm{u_{V}}u_\mathrm{V}, \quad \phi_\mathrm{C_{leak}} = (C_\mathrm{out}-C)k_\mathrm{leak},
\end{equation}
where $k_\mathrm{leak}$ is the leakage factor.
%where $C_\mathrm{out}$ is the concentration of $CO_2$ of the external source. The outflux of the molecules through leakage vents due to gradient in gas concentration can be expressed as
%\begin{equation}\label{eq:phi_cout}
%\phi_\mathrm{C_{leak}} = (C_\mathrm{out}-C)k_\mathrm{leak},
%\end{equation}
%where $k_\mathrm{leak}$ is the leakage factor.
The dynamics of the $CO_2$ concentration can be derived from the mass balance equation as
\begin{equation}\label{eq:dC/dt}
%k_\mathrm{V,chm}\frac{\mathrm{d}C}{\mathrm{d}t}= \phi_\mathrm{C_{exch}}+\phi_\mathrm{C_{leak}}+ \phi_\mathrm{C_{sub}},
k_\mathrm{V,chm}\dot{C}= \phi_\mathrm{C_{exch}}+\phi_\mathrm{C_{leak}}+ \phi_\mathrm{C_{sub}},
\end{equation}
where $\phi_\mathrm{C_{sub}}$ is the net flux contributed by the metabolic activities of the subject and $k_\mathrm{V,chm}$ is the volume inside the growing chamber. %Similarly, the equation for $O_2$ is
%\begin{equation}\label{eq:dO/dt}
%V_\mathrm{chm}\frac{\mathrm{d}O}{\mathrm{d}t}= %\phi_\mathrm{O_{exch}}-\phi_\mathrm{O_{leak}}+ \phi_\mathrm{O_{sub}}.
%\end{equation}

\subsubsection{Water Flux}
Water flux within the chamber and to the outside occurs in both gaseous and liquid forms. In gaseous form, water is mixed in the air and contributes to the humidity.
The change in humidity due to the air exchange with external source and the ultrasonic humidifier can be defined respectively as
\begin{equation}
\phi_\mathrm{H_{exch}} = (H_\mathrm{out} - H)k_\mathrm{u_{V}}u_\mathrm{V},\quad \phi_\mathrm{u_H} = (H_\mathrm{sat}(T) - H)k_\mathrm{u_{H}}u_\mathrm{H},
\end{equation}
where $k_\mathrm{u_{H}}$ is the humidification rate.
The saturation concentration of water vapor $H_\mathrm{sat}$  for a reference temperature $T_\mathrm{ref}$, can be calculated using the Magnus-Tetens equation \citep{Murray1967} as
\begin{equation}
H_\mathrm{sat} = \frac{k_\mathrm{mw}}{k_\mathrm{R,g}(T_\mathrm{ref}+273)}
\left(0.61094\cdot e^{\left(\frac{17.625\cdot T_\mathrm{ref}}{T_\mathrm{ref}+243.03}\right)}\right),
\end{equation}
where $k_\mathrm{mw}$ is the molar mass of water and $k_\mathrm{R,g}$ is the gas constant.

%When the temperature of the heat exchanging surface of the condenser falls below the dew-point temperature, water vapor in contact with this surface starts to condense. This condensed water, collected in the storage tank, can be modeled as a function of saturation concentration of water vapor at the surface temperature of the condenser $H_\mathrm{sat}$ and the surface area of the condenser $k_\mathrm{a,cond}$ \citep{Bot1983,STRATEN}
Condensation of water on the heat exchanging surface can be modeled as a function of saturation concentration of water vapor $H_\mathrm{sat}$ at the surface temperature of the condenser $T_\mathrm{c}$ and the surface area of the condenser $k_\mathrm{a,cond}$ \citep{STRATEN2011} as
\begin{equation}\label{eq:phi_cond}
\phi_\mathrm{W_{cond}} = \max\left( \frac{k_\mathrm{a,cond} k_\mathrm{h,cond}}{k_\mathrm{\rho}k_\mathrm{c}{k_\mathrm{Le}}^{\frac{2}{3}}}
\left(H - H_\mathrm{sat}\left(T_\mathrm{c}\right)\right),0\right),
\end{equation}
where $k_\mathrm{h,cond}$ is the heat transfer coefficient and $k_\mathrm{Le}$ is the Lewis number for water vapor.% and the temperature on the surface of the condenser $T_\mathrm{cond}= T + \left(\frac{\phi_{\mathrm{Q_{TEC}}}}{k_\mathrm{C,cond}}\right)$ with $k_\mathrm{C,cond}$ being the heat capacity of the condenser.

The final equation describing the humidity flux can be summarized as:
\begin{equation}\label{eq:dH/dt}
%{k_\mathrm{V,cham}}\frac{\mathrm{d}H}{\mathrm{d}t}= \phi_\mathrm{H_{exch}} + \phi_\mathrm{u_{H}} - \phi_\mathrm{W_{cond}} +\phi_\mathrm{H_{sub}},%+\phi_\mathrm{W_{evap}}
{k_\mathrm{V,cham}}\dot{H}= \phi_\mathrm{H_{exch}} + \phi_\mathrm{u_{H}} - \phi_\mathrm{W_{cond}} +\phi_\mathrm{H_{sub}},%+\phi_\mathrm{W_{evap}}
\end{equation}
where $\phi_\mathrm{H_{sub}}$ is the transpiration from subject. %, and $\phi_\mathrm{W_{evap}}$ is the rate of evaporation from the surface of the growing medium.

Water influx to the chamber in liquid form occurs from two different sources (M3, M4). %Water that enters the chamber from motor M4 is stored in the growing medium while the water from M3 is stored in the internal storage tank.
These fluxes are modeled using state variables: $W_\mathrm{sto}$, water in internal storage tank; $W_\mathrm{med}$, water in the growing medium; and $W_\mathrm{ovf}$, water overflowing from both the storage tank and growing medium. These dynamics are modeled as
\begin{equation}
\dot{W_\mathrm{sto}} = k_\mathrm{u_W}u_\mathrm{W1}+\phi_\mathrm{W_{cond}}-\phi_\mathrm{u_H}-\phi_\mathrm{W_{ovf}1}, \label{eq:dW_sto/dt}
\end{equation}
\begin{align}
\dot{W_\mathrm{med}} &= k_\mathrm{u_W}u_\mathrm{W2}-\phi_\mathrm{W_{evap}}-\phi_\mathrm{W_{sub}}-\phi_\mathrm{W_{ovf}2}, \label{eq:dW_med/dt}\\
\dot{W_\mathrm{ovf}} &= \phi_\mathrm{W_{ovf}1} + \phi_\mathrm{W_{ovf}2}  -k_\mathrm{u_W}u_\mathrm{W3},\label{eq:dW_ovf/dt}
\end{align}
where $\phi_\mathrm{W_{1}}$ and $\phi_\mathrm{W_{2}}$ are water pumped into the storage tank and the growing medium respectively, and $\phi_\mathrm{W_{sub}}$ is the water consumed by the subject.
$\phi_\mathrm{W_{ovf}1}$ and $\phi_\mathrm{W_{ovf}2}$ represents the water that overflows from the storage tank and the growth medium and $k_\mathrm{u_W}$ is the output rate of the water pumps.

The flux terms on the right hand side of the equation \eqref{eq:dW_sto/dt} and \eqref{eq:dW_med/dt} excluding the overflow terms represents the effective water flow, $\phi_\mathrm{W_{eff}}$, into the respective containers.
Since the overflow occurs only when the tank reaches its maximum capacity, $k_\mathrm{Wm}$, this overflow is modeled as
\begin{align}
	\begin{split}
	\phi_\mathrm{W_{ovf}}
	&= \begin{cases} 
	0       					& \text{if } W_\mathrm{sto} \leq k_\mathrm{Wm} \\
	\phi_\mathrm{W_{eff}}       & \text{if } W_\mathrm{sto} > k_\mathrm{Wm}
	\end{cases}.
	\label{eq:phi_Wo}
	\end{split}
\end{align}

%\subsubsection{Radiant Flux}
%The total radiant flux contributed by the LED light panel is given as a sum of usable spectral irradiance of individual LED channels 
%\begin{equation}\label{eq:dI/dt}
%I = \sum_{i=1}^{4}\eta_{\mathrm{LU}i}u_{\mathrm{I}i}\phi_{\mathrm{I}i},
%\end{equation}
%where $\eta_\mathrm{LU}$ is the light utilization efficiency specific to the subject growing inside. \red{ What is $\phi_{\mathrm{I}i}$? should be may be replaced with maximum intensity/PAR etc.,}

\subsection{Plant model}\label{sec:plant_model}
 A dynamic growth model of Lettuce presented in \citep{STRATEN2011}, is used as the subject of interest growing in the chamber. This model considers the effect of light $I$, temperature $T$ and CO$_2$ concentration $C$ on the lettuce growth, ignoring the effects of day-night light cycles. The two major resource dynamics addressed in this model are the CO$_2$ and water which is consumed and converted into plant dry weight $B$ (biomass) normalized to the available area of the growing medium $k_\mathrm{a,med}$.

The net CO$_2$ change rate $\phi_\mathrm{C_{sub}}$ due to photosynthesis and respiration is given as sum of the following two components
\begin{align}
	\begin{split}
	\phi_\mathrm{C_{phot}}
	&= k_\mathrm{a,med}\left(1-e^{-k_{\mathrm{LAI}}B}\right)\\
	&\left(\frac{k_\mathrm{I,p}I(-k_\mathrm{p,1}T^2 + k_\mathrm{p,2}T -	k_\mathrm{p,3})(C - k_\mathrm{\Gamma,p})}{k_\mathrm{I,p}I + (-k_\mathrm{p,1}T^2 + k_\mathrm{p,2}T - k_\mathrm{p,3})(C - k_\mathrm{\Gamma,p})}
	\right),
	\label{eq:phi_Cin}
	\end{split}
\end{align}
\begin{equation}
\phi_\mathrm{C_{resp}}
=k_\mathrm{a,med}k_\mathrm{resp}B\cdot 2^{(0.1T - 2.5)},
\label{eq:phi_Cout}
\end{equation}
where $k_\mathrm{resp}$ is the respiration coefficient, $k_{\mathrm{LAI}}$ is the effective canopy area per kilogram of biomass, $k_\mathrm{I,p}$ is the light utilization efficiency of plant, $k_\mathrm{p,1}$, $k_\mathrm{p,2}$ and $k_\mathrm{p,3}$ are empirically derived parameters for temperature dependence, and $k_\mathrm{\Gamma,p}$ is the CO$_2$ compensation point.

Humidity and water flux components contributed by the plant can be summarized respectively as
\begin{align}
\phi_\mathrm{H_{sub}}
&= k_\mathrm{a,med}k_\mathrm{H,trans} \left(1-e^{-k_{\mathrm{LAI}}B}\right) \left(H_\mathrm{sat} - H\right),\\
\phi_{\mathrm{W_{sub}}}
&= \phi_\mathrm{H_{sub}} + k_\mathrm{a,med}(1-k_\mathrm{fw,dw})\dot{B},
\label{eq:phi_W_evap}
\end{align}
where $k_\mathrm{H,trans}$ is the mass transfer coefficient and $k_\mathrm{fw,dw}$ is the plant fresh to dry weight ratio.

Equations \eqref{eq:phi_Cout} and \eqref{eq:phi_Cin} are used to model the biomass rate change as
\begin{equation}\label{eq:dB/dt}
\dot{B}
=k_\mathrm{\alpha,\beta} \phi_\mathrm{C_{phot}}
- k_\mathrm{B_{resp}}\phi_\mathrm{C_{resp}},
\end{equation}
where $k_\mathrm{\alpha,\beta}$ is the biomass conversion per kilogram of CO$_2$ assimilated and $k_\mathrm{B_{resp}}$ is the respiration rate.

\subsection{Combined system}

Equations governing the system states and the resource flux terms presented in \eqref{eq:dT/dt}, \eqref{eq:dC/dt}, \eqref{eq:dH/dt}, \eqref{eq:dW_sto/dt}, \eqref{eq:dW_med/dt}, \eqref{eq:dW_ovf/dt} and \eqref{eq:dB/dt} summarize the system under consideration.
This system under study can be of the form:
\begin{align}\label{eq:sys}
\mathbf{\dot{x}} &= \mathbf{f}(\mathbf{x}, \mathbf{u}, \mathbf{d}),
%\mathbf{y} &= \mathbf{x},
\end{align}
with the state vector $\mathbf{x}=\mathbf{x}(t)$, input vector $\mathbf{u}=\mathbf{u}(t)$ and disturbance vector $\mathbf{d}=\mathbf{d}(t)$ given as
%\begin{equation*}
%\mathbf{x}= 
%\begin{bmatrix} T \\
%C\\
%H\\
%W_\mathrm{sto}\\
%W_\mathrm{med}\\
%W_\mathrm{ovf}\\
%B\\
%\end{bmatrix}, \;
%\mathbf{u}= 
%\begin{bmatrix} u_\mathrm{T} \\
%u_\mathrm{V}\\
%u_\mathrm{H}\\
%u_\mathrm{W1}\\
%u_\mathrm{W2}\\
%u_\mathrm{W3}\\
%u_{\mathrm{I}1}\\
%u_{\mathrm{I}2}\\
%u_{\mathrm{I}3}\\
%u_{\mathrm{I}4}\\
%\end{bmatrix}, \;
%\mathbf{d}= 
%\begin{bmatrix} T_\mathrm{out} \\
%C_\mathrm{out}\\
%H_\mathrm{out}\\
%%I_\mathrm{out}\\
%\end{bmatrix}. 
%\end{equation*}
\begin{align*}
\mathbf{x} &= 
\begin{bmatrix} T &
C&
H&
W_\mathrm{sto}&
W_\mathrm{med}&
W_\mathrm{ovf}&
B
\end{bmatrix}^\top, \\
\mathbf{d} &= 
\begin{bmatrix} T_\mathrm{out} &
C_\mathrm{out} &
H_\mathrm{out} 
%I_\mathrm{out}\\
\end{bmatrix}^\top, 
\\
\mathbf{u} &= 
\begin{bmatrix} u_\mathrm{T} &
u_\mathrm{V}&
u_\mathrm{H}&
u_\mathrm{W1}&
u_\mathrm{W2}&
u_\mathrm{W3}&
u_{\mathrm{I}1}&
u_{\mathrm{I}2}&
u_{\mathrm{I}3}&
u_{\mathrm{I}4}
\end{bmatrix}^\top.
\end{align*}
Constraints of the actuators, due to their construction, are
\begin{align*}
u_\mathrm{T} \in[-100,100], \; u_{\mathrm{I}i} &\in[0,100],\\
u_\mathrm{V},u_\mathrm{H},u_\mathrm{W1},u_\mathrm{W2},u_\mathrm{W3} &\in\{0,1\},
\end{align*}
which for brevity are referred to as $\mathbb{U} \subset \mathbb{R}^{10}$.

Constraints of the states are given by
\begin{align*}
T &\in[5,40], \; C \in[\num{1.96e-06}, \num{1.7e-2}],\\
H &\in[\num{4.85e-05}, \num{5.1e-2}], \; W_\mathrm{sto} \in[\num{1e-04}, 0.3],\\
W_\mathrm{med} &\in[0.3, 1], \;
W_\mathrm{ovf} \in[0.1, 2], \;
B \in[\num{1e-6}, 0.5],
\end{align*}
for brevity comprised to $\mathbb{X} \subset \mathbb{R}^{7}$, that include plant's survival conditions as well as the production unit specification (e.g. size of water tanks).

%% file: gfx/Plant_process_growth_chamber.pdf_tex
%% Accompanies image file 'Plant_process_growth_chamber.pdf' (pdf, eps, ps)
\begingroup%
  \makeatletter%
  \providecommand\color[2][]{%
    \errmessage{(Inkscape) Color is used for the text in Inkscape, but the package 'color.sty' is not loaded}%
    \renewcommand\color[2][]{}%
  }%
  \providecommand\transparent[1]{%
    \errmessage{(Inkscape) Transparency is used (non-zero) for the text in Inkscape, but the package 'transparent.sty' is not loaded}%
    \renewcommand\transparent[1]{}%
  }%
  \providecommand\rotatebox[2]{#2}%
  \newcommand*\fsize{\dimexpr\f@size pt\relax}%
  \newcommand*\lineheight[1]{\fontsize{\fsize}{#1\fsize}\selectfont}%
  \ifx\svgwidth\undefined%
    \setlength{\unitlength}{1355.22021516bp}%
    \ifx\svgscale\undefined%
      \relax%
    \else%
      \setlength{\unitlength}{\unitlength * \real{\svgscale}}%
    \fi%
  \else%
    \setlength{\unitlength}{\svgwidth}%
  \fi%
  \global\let\svgwidth\undefined%
  \global\let\svgscale\undefined%
  \makeatother%
  \begin{picture}(1,0.51465818)%
    \lineheight{1}%
    \setlength\tabcolsep{0pt}%
    \put(0,0){\includegraphics[width=\unitlength,page=1]{Plant_process_growth_chamber.pdf}}%
    \put(0.52127856,0.28669643){\color[rgb]{0,0,0}\makebox(0,0)[lt]{\lineheight{1.25}\smash{\begin{tabular}[t]{l}\textbf{$\phi_\mathrm{Q_{exch}}$}\end{tabular}}}}%
    \put(0,0){\includegraphics[width=\unitlength,page=2]{Plant_process_growth_chamber.pdf}}%
    \put(0.39450433,0.28738163){\color[rgb]{0,0,0}\makebox(0,0)[lt]{\lineheight{1.25}\smash{\begin{tabular}[t]{l}\textbf{$\phi_\mathrm{H_{exch}}$}\end{tabular}}}}%
    \put(0,0){\includegraphics[width=\unitlength,page=3]{Plant_process_growth_chamber.pdf}}%
    \put(0.49780263,0.24899379){\color[rgb]{0,0,0}\makebox(0,0)[lt]{\lineheight{1.25}\smash{\begin{tabular}[t]{l}\textbf{$\phi_\mathrm{C_{exch}}$}\end{tabular}}}}%
    \put(0,0){\includegraphics[width=\unitlength,page=4]{Plant_process_growth_chamber.pdf}}%
    \put(0.42268933,0.49342766){\color[rgb]{0,0,0}\makebox(0,0)[lt]{\lineheight{1.25}\smash{\begin{tabular}[t]{l}Chamber	unit\end{tabular}}}}%
    \put(0.29979066,0.34499602){\color[rgb]{0,0,0}\makebox(0,0)[lt]{\lineheight{1.25}\smash{\begin{tabular}[t]{l}Air	\end{tabular}}}}%
    \put(0.28070962,0.32483742){\color[rgb]{0,0,0}\makebox(0,0)[lt]{\lineheight{1.25}\smash{\begin{tabular}[t]{l}circulator\end{tabular}}}}%
    \put(0.1995447,0.25799574){\color[rgb]{0,0,0}\makebox(0,0)[lt]{\lineheight{1.25}\smash{\begin{tabular}[t]{l}Condensor\end{tabular}}}}%
    \put(0.14012988,0.35348387){\color[rgb]{0,0,0}\makebox(0,0)[lt]{\lineheight{1.25}\smash{\begin{tabular}[t]{l}TEC\end{tabular}}}}%
    \put(0.06266505,0.20600778){\color[rgb]{0,0,0}\makebox(0,0)[lt]{\lineheight{1.25}\smash{\begin{tabular}[t]{l}M3\end{tabular}}}}%
    \put(0.06266505,0.12749532){\color[rgb]{0,0,0}\makebox(0,0)[lt]{\lineheight{1.25}\smash{\begin{tabular}[t]{l}M4\end{tabular}}}}%
    \put(0.87582084,0.2144956){\color[rgb]{0,0,0}\makebox(0,0)[lt]{\lineheight{1.25}\smash{\begin{tabular}[t]{l}M2\end{tabular}}}}%
    \put(0.90468972,0.04815115){\color[rgb]{0,0,0}\makebox(0,0)[lt]{\lineheight{1.25}\smash{\begin{tabular}[t]{l}Controller\end{tabular}}}}%
    \put(0.26214248,0.17523936){\color[rgb]{0,0,0}\makebox(0,0)[lt]{\lineheight{1.25}\smash{\begin{tabular}[t]{l}Humidifier\end{tabular}}}}%
    \put(0.65555019,0.11423308){\color[rgb]{0,0,0}\makebox(0,0)[lt]{\lineheight{1.25}\smash{\begin{tabular}[t]{l}S4		\end{tabular}}}}%
    \put(0.67570882,0.44414535){\color[rgb]{0,0,0}\makebox(0,0)[lt]{\lineheight{1.25}\smash{\begin{tabular}[t]{l}S1\end{tabular}}}}%
    \put(0.67623927,0.40435863){\color[rgb]{0,0,0}\makebox(0,0)[lt]{\lineheight{1.25}\smash{\begin{tabular}[t]{l}S2\end{tabular}}}}%
    \put(0.06242974,0.34552654){\color[rgb]{0,0,0}\makebox(0,0)[lt]{\lineheight{1.25}\smash{\begin{tabular}[t]{l}Heat	\end{tabular}}}}%
    \put(0.04570274,0.32536793){\color[rgb]{0,0,0}\makebox(0,0)[lt]{\lineheight{1.25}\smash{\begin{tabular}[t]{l}exchanger\end{tabular}}}}%
    \put(0.67570882,0.36298044){\color[rgb]{0,0,0}\makebox(0,0)[lt]{\lineheight{1.25}\smash{\begin{tabular}[t]{l}S3\end{tabular}}}}%
    \put(0.87688174,0.12112948){\color[rgb]{0,0,0}\makebox(0,0)[lt]{\lineheight{1.25}\smash{\begin{tabular}[t]{l}M1\end{tabular}}}}%
    \put(0.065848,0.04845238){\color[rgb]{0,0,0}\makebox(0,0)[lt]{\lineheight{1.25}\smash{\begin{tabular}[t]{l}M5\end{tabular}}}}%
    \put(0.21839366,0.34817896){\color[rgb]{0,0,0}\makebox(0,0)[lt]{\lineheight{1.25}\smash{\begin{tabular}[t]{l}Heat	\end{tabular}}}}%
    \put(0.20166666,0.32802036){\color[rgb]{0,0,0}\makebox(0,0)[lt]{\lineheight{1.25}\smash{\begin{tabular}[t]{l}exchanger\end{tabular}}}}%
    \put(0.86053471,0.43836219){\color[rgb]{0,0,0}\makebox(0,0)[lt]{\lineheight{1.25}\smash{\begin{tabular}[t]{l}Heat	Energy	Flow\end{tabular}}}}%
    \put(0.86053471,0.41714262){\color[rgb]{0,0,0}\makebox(0,0)[lt]{\lineheight{1.25}\smash{\begin{tabular}[t]{l}Water	Flow\end{tabular}}}}%
    \put(0.86053471,0.39592303){\color[rgb]{0,0,0}\makebox(0,0)[lt]{\lineheight{1.25}\smash{\begin{tabular}[t]{l}Air	Flow\end{tabular}}}}%
    \put(0.86053471,0.37470345){\color[rgb]{0,0,0}\makebox(0,0)[lt]{\lineheight{1.25}\smash{\begin{tabular}[t]{l}Data	Flow\end{tabular}}}}%
    \put(0.86106522,0.35613632){\color[rgb]{0,0,0}\makebox(0,0)[lt]{\lineheight{1.25}\smash{\begin{tabular}[t]{l}Control	Signal/\end{tabular}}}}%
    \put(0.86106522,0.33597772){\color[rgb]{0,0,0}\makebox(0,0)[lt]{\lineheight{1.25}\smash{\begin{tabular}[t]{l}Electric	Energy	Flow\end{tabular}}}}%
    \put(0.65289778,0.06755002){\color[rgb]{0,0,0}\makebox(0,0)[lt]{\lineheight{1.25}\smash{\begin{tabular}[t]{l}S5				\end{tabular}}}}%
    \put(0.00957143,0.47549647){\color[rgb]{0,0,0}\makebox(0,0)[lt]{\lineheight{1.25}\smash{\begin{tabular}[t]{l}Air	conditioning	unit\end{tabular}}}}%
    \put(0.47106324,0.0486946){\color[rgb]{0,0,0}\makebox(0,0)[lt]{\lineheight{1.25}\smash{\begin{tabular}[t]{l}\textbf{$\phi_\mathrm{W_{sub}}$}\end{tabular}}}}%
    \put(0.22034091,0.19872339){\color[rgb]{0,0,0}\makebox(0,0)[lt]{\lineheight{1.25}\smash{\begin{tabular}[t]{l}\textbf{$W_\mathrm{sto}$}\end{tabular}}}}%
    \put(0.4785612,0.14522356){\color[rgb]{0,0,0}\makebox(0,0)[lt]{\lineheight{1.25}\smash{\begin{tabular}[t]{l}$B$\end{tabular}}}}%
    \put(0.48248524,0.21425182){\color[rgb]{0,0,0}\makebox(0,0)[lt]{\lineheight{1.25}\smash{\begin{tabular}[t]{l}\textbf{$\phi_\mathrm{C_{sub}}$}\end{tabular}}}}%
    \put(0.39110657,0.17610378){\color[rgb]{0,0,0}\makebox(0,0)[lt]{\lineheight{1.25}\smash{\begin{tabular}[t]{l}\textbf{$\phi_\mathrm{H_{sub}}$}\end{tabular}}}}%
    \put(0.25284393,0.27543255){\color[rgb]{0,0,0}\makebox(0,0)[lt]{\lineheight{1.25}\smash{\begin{tabular}[t]{l}\textbf{$\phi_\mathrm{W_{TEC}}$}\end{tabular}}}}%
    \put(0.38289391,0.38816625){\color[rgb]{0,0,0}\makebox(0,0)[lt]{\lineheight{1.25}\smash{\begin{tabular}[t]{l}\textbf{$\phi_\mathrm{Q_{TEC}}$}\end{tabular}}}}%
    \put(0.46760199,0.40865922){\color[rgb]{0,0,0}\makebox(0,0)[lt]{\lineheight{1.25}\smash{\begin{tabular}[t]{l}\textbf{$\phi_\mathrm{I_{LED}}$}\end{tabular}}}}%
    \put(0.46105028,0.3602986){\color[rgb]{0,0,0}\makebox(0,0)[lt]{\lineheight{1.25}\smash{\begin{tabular}[t]{l}\textbf{$T,C,H$}\end{tabular}}}}%
    \put(0.91601961,0.18879606){\color[rgb]{0,0,0}\makebox(0,0)[lt]{\lineheight{1.25}\smash{\begin{tabular}[t]{l}\textbf{$T_\mathrm{out}$ }\\\textbf{$C_\mathrm{out}$}\\\textbf{$H_\mathrm{out}$}\end{tabular}}}}%
    \put(0.36092641,0.04913864){\color[rgb]{0,0,0}\makebox(0,0)[lt]{\lineheight{1.25}\smash{\begin{tabular}[t]{l}\textbf{$W_\mathrm{med}$}\end{tabular}}}}%
    \put(0.23533488,0.03096905){\color[rgb]{0,0,0}\makebox(0,0)[lt]{\lineheight{1.25}\smash{\begin{tabular}[t]{l}\textbf{$W_\mathrm{ovf}$}\end{tabular}}}}%
    \put(0.11223928,0.13558432){\color[rgb]{0,0,0}\makebox(0,0)[lt]{\lineheight{1.25}\smash{\begin{tabular}[t]{l}\textbf{$k_\mathrm{u_W}u_\mathrm{W2}$}\end{tabular}}}}%
    \put(0.11179735,0.06359074){\color[rgb]{0,0,0}\makebox(0,0)[lt]{\lineheight{1.25}\smash{\begin{tabular}[t]{l}\textbf{$k_\mathrm{u_W}u_\mathrm{W3}$}\end{tabular}}}}%
    \put(0.10898357,0.21914089){\color[rgb]{0,0,0}\makebox(0,0)[lt]{\lineheight{1.25}\smash{\begin{tabular}[t]{l}\textbf{$k_\mathrm{u_W}u_\mathrm{W1}$}\end{tabular}}}}%
    \put(0.22735913,0.11637141){\color[rgb]{0,0,0}\makebox(0,0)[lt]{\lineheight{1.25}\smash{\begin{tabular}[t]{l}\textbf{$\phi_\mathrm{W_{ovf}1}$}\end{tabular}}}}%
    \put(0.27233929,0.2316934){\color[rgb]{0,0,0}\makebox(0,0)[lt]{\lineheight{1.25}\smash{\begin{tabular}[t]{l}\textbf{$k_\mathrm{u_H}u_\mathrm{H}$}\end{tabular}}}}%
    \put(0.25450824,0.07985652){\color[rgb]{0,0,0}\makebox(0,0)[lt]{\lineheight{1.25}\smash{\begin{tabular}[t]{l}\textbf{$\phi_\mathrm{W_{ovf}2}$}\end{tabular}}}}%
    \put(0.41945544,0.44617005){\color[rgb]{0,0,0}\makebox(0,0)[lt]{\lineheight{1.25}\smash{\begin{tabular}[t]{l}\textbf{$\phi_\mathrm{Q_{LED}}$}\end{tabular}}}}%
    \put(0.79312946,0.22891463){\color[rgb]{0,0,0}\makebox(0,0)[lt]{\lineheight{1.25}\smash{\begin{tabular}[t]{l}\textbf{$k_\mathrm{u_V}u_\mathrm{V}$}\end{tabular}}}}%
    \put(0.79459461,0.14045981){\color[rgb]{0,0,0}\makebox(0,0)[lt]{\lineheight{1.25}\smash{\begin{tabular}[t]{l}\textbf{$k_\mathrm{u_V}u_\mathrm{V}$}\end{tabular}}}}%
    \put(0.52959724,0.17637403){\color[rgb]{0,0,0}\makebox(0,0)[lt]{\lineheight{1.25}\smash{\begin{tabular}[t]{l}\textbf{$\phi_\mathrm{Q_{sub}}$}\end{tabular}}}}%
    \put(0,0){\includegraphics[width=\unitlength,page=5]{Plant_process_growth_chamber.pdf}}%
    \put(0.80224974,0.26932603){\color[rgb]{0,0,0}\makebox(0,0)[lt]{\lineheight{1.25}\smash{\begin{tabular}[t]{l}\textbf{$\phi_\mathrm{Q_{loss}}$}\end{tabular}}}}%
    \put(0,0){\includegraphics[width=\unitlength,page=6]{Plant_process_growth_chamber.pdf}}%
    \put(0.80296164,0.30887035){\color[rgb]{0,0,0}\makebox(0,0)[lt]{\lineheight{1.25}\smash{\begin{tabular}[t]{l}\textbf{$\phi_\mathrm{C_{leak}}$}\end{tabular}}}}%
    \put(0,0){\includegraphics[width=\unitlength,page=7]{Plant_process_growth_chamber.pdf}}%
    \put(0.35404303,0.23879154){\color[rgb]{0,0,0}\makebox(0,0)[lt]{\lineheight{1.25}\smash{\begin{tabular}[t]{l}\textbf{$\phi_\mathrm{u_{H}}$}\end{tabular}}}}%
  \end{picture}%
\endgroup%

%% file: control.tex
\section{Production Unit Control}\label{sec:control}

In this section, optimal control of the production unit is addressed with particular focus on the temperature and CO$_2$ levels. 
%\red{Since the entire state can be measured, state feedback methods could be employed. }
%This is particularly suitable in case all states are to follow a reference trajectory. 
Because binary constraints are present, and the overall objective is to not violate growth constraints while minimizing the energy demand, a predictive controller with relaxed constraint specification is employed.
The task of state estimation is omitted due to page limitation and full state availability at every time instance is assumed. 
As to this point, control is designed as a tracking problem of particular desired temperature and CO$_2$ levels whereas economic factors embodied in the optimization objective (refer to economic MPC, cf. \cite{Rawlings2012}) may be considered as well.

\subsection{System Discretization and Predictive Control Design}
Predictive control is performed at discrete time step, for which the dynamics $x(t_k + 1) = x(t_k) + \Delta t f(x(t_k),u(t_k),d(t_k)) =: f_d(x(t_k),u(t_k),d(t_k))$ are utilized, where $t_k = n \Delta t$, $n \in \N_0$, and $\Delta t >0$ is the sample time. 
Due to the ``high'' nonlinearity of the continuous-time dynamics, a sufficiently small sample time should be chosen such that the behaviour is approximated adequately on a given time interval \eg the prediction horizon. 
However, using simulations of the system response to sample controls for a set of initial states, it could be observed that the state changes are relatively slow compared to the timescale of the system \sut a model discretization time of $\Delta t = 30$ sec is regarded sufficient for control. 
The predictive controller is applied on the nonlinear model, under awareness of the numerical difficulties involved in nonlinear optimization \citep[see \eg][for a study]{Kamel2017}, with a sampling frequency of $\Delta t = 30$ sec, which equals that of the model discretization. 
Using the prediction horizon $N=5$, which yields a $2.5$ min lookahead time, the optimization setup is rendered sufficiently fast (computationally).%, and avoids other methods for acceleration of nonlinear optimization such as \eg linearization or move blocking \citep[see \eg][]{Cagienard2007}. 
%For a discussion on the computational difficulties of linear and nonlinear MPC tracking performance refer to \eg \citep{Kamel2017}.
The particular difficulty in using longer prediction horizons lies in the fact that the computational load increases significantly. 
Although local linearizations could be used to reduce this burden, linear dynamics approximation may be unsuitable when predicting over longer horizons in which the state and/or input may reach values ``outside'' the validity of the linear approximation. 
It will be shown in Section \ref{sec:result}, however, that the horizon length is sufficient for constraint satisfaction as well as efficient reference tracking. 

%\red{\st{At this point, the switching nature of the function $\phi_{Wovf}$ in} \eqref{eq:phi_Wo} \st{need to be discussed.
%It can be observed that switching occurs at values equivalent to the lower bound of the state constraints to $W_{\text{med}}$ and $W_{\text{sto}}$. 
%Therefore, restriction to one case can be justified by operating the system above the lower bound, using respective state constraints in the optimization.}}

%For optimization, the horizon length $N=5$ under $\Delta t = 30$ sec is used, which equals a prediction horizon of $2.5$ min. 
The (bounded) disturbance $\mathbf{d}(t)$ is assumed to be known for the entire horizon length $[t_k,t_k + N]$ for any $t_k$ \eg by using short-term weather forecast, while at each time step, the discretized trajectory is shifted and only the last value at $t_k + N$ is updated. 
That is, at each time instance, the previous climate data for the specified time horizon remains as predicted while adding a new measure $\mathbf{d}(t_k+N)$ to the sequence.
This allows to consider the time-varying system $f_d(\mathbf{x}(t_k),\mathbf{u}(t_k),\mathbf{d}(t_k)) = f_d(\mathbf{x}(t_k),\mathbf{u}(t_k),t_k)$. 
As this could be considered a strong assumption, it should be pointed out that short-term weather forecast supplies reliable and sufficiently accurate data. Whereas the external disturbance may additionally be regarded as near constant on the inspected time interval, given a current environment state. 

%Define the weighted norm $\|\mathbf{x}\|_P^2 = \mathbf{x}^\top P \mathbf{x}$, where $P$ is a positive semi-definite
At every time $t_k$, the solution to
\begin{subequations} \label{eq:MPC}
	\begin{align}
	\begin{split}
	\min_{\substack{\varepsilon,\mathbf{u}_{i}(t_k) \\ i=0,\dotsc,N-1}} &\sum_{i=0}^{N-1} r(\mathbf{x}_{i}(t_k) - \mathbf{x}^{\text{ref}}(t_k +i), \mathbf{u}_{i}(t_k)-\mathbf{u}^{\text{ref}}(t_k+i)   ) \\
	%&\hspace{1cm}+\|\mathbf{x}_{N}(t_k) - \mathbf{x}^{\text{ref}}(t_k +N)\|_P^2  + \alpha \varepsilon^2   \label{eq:MPC-1}
	&+ V_P(\mathbf{x}_{N}(t_k))  + \alpha \varepsilon^2   \label{eq:MPC-1}
	\end{split}\\
	\text{s.t.} \quad &\mathbf{x}_{i+1}(t_k) = f_d(\mathbf{x}_{i}(t_k),\mathbf{u}_{i}(t_k),t_k+i) \\
	&\mathbf{x}_{0}(t_k) = \mathbf{x}(t_k) \\
	& \mathbf{x}_{i}(t_k) \in \X, \; i = 0,\dotsc,N \\
	& \mathbf{u}_{i}(t_k) \in \U, \; i = 0,\dotsc,N-1 \\ 
	& \nu_{i}(t_k) = \nu_{i}^2(t_k) + \varepsilon, \; i = 0,\dotsc,N-1 \label{eq:MPC-6} 
	\end{align}
\end{subequations}
is computed for some $\alpha \gg 1$ and $V_P(x)= \left( x - \mathbf{x}^{\text{ref}}(t_k +N)\right)^\top P\left( x - \mathbf{x}^{\text{ref}}(t_k +N)\right)$, with $P \succeq 0$.
The optimization \eqref{eq:MPC} yields the minimizing control sequence $\{\mathbf{u}_{0}^\ast(t_k),\dotsc,\mathbf{u}_{N-1}^\ast(t_k)\}$, while $\mathbf{u}_0^\ast(t_k) =: \mathbf{u}^\ast(t_k)$ is applied to the system, as well as an optimal relaxation $\varepsilon^\ast(t_k)$. 
The binary constraints in $\mathbb{U}$ are tackled via constraint relaxation according to \eqref{eq:MPC-6}. 
Forcing $\varepsilon \rightarrow 0$ through the cost $\alpha \varepsilon^2$ with $\alpha \gg 1$ renders $\nu \in \{0,1\}$ the only admissible values, whereas \eqref{eq:MPC-6} comprises a set of constraints for all $\{u_{\text{V}}, u_{\text{H}},u_{\text{W}1},u_{\text{W2}},u_{\text{W}3}\} \ni \nu$.  
%The constraint \eqref{eq:MPC-6} seeks to relax the binary constraints by using $\nu = \{u_{\text{V}}, u_{\text{H}},u_{\text{W}1},u_{\text{W2}},u_{\text{W}3}\}$ %\blue{and forcing $\varepsilon \rightarrow 0$ by adding $\alpha \varepsilon^2$ with $\alpha \gg 1$}. % and $\varepsilon \ra 0$, $\nu = \nu^2$ admits $\nu=0$ or $\nu=1$ as 
%The last term of the objective intends to penalize the relaxation variable $\varepsilon$ on the binary constraints \eqref{eq:MPC-6}, since for $\varepsilon = %0$, $\nu \in \{0,1\}$ are admissible only.
In \eqref{eq:MPC-1}, $\mathbf{x}^{\text{ref}}(t),\, \mathbf{u}^{\text{ref}}(t)$ are reference trajectories to be specified in the following Section \ref{subsec:tracking}, $r(x,u):\mathbb{X} \times \mathbb{U} \rightarrow \mathbb{R}_{\geq 0}$ is a positive semi-definite running cost and $P = \text{diag}(5000,\num{1.1e12},0,\dotsc,0)$ is a terminal weight matrix. 
%Notice that no terminal constraint is imposed on the state $x_N(t_k)$, as such was observed to lead to feasibility issues -- these may occur whenever the disturbance cannot be compensated by the actuators due to their physical limitations. 
%In these cases, no horizon length can be found such that the terminal state can be driven to the desired value under the given actuator specifications.

%Additionally, such constraint lead to 

\subsection{Reference Specification}
\label{subsec:tracking}

 Plant growth can be quantified primarily over the instantaneous photosynthetic rate (occurring at $s^{-1}$ rate) and the net assimilation over 24 hrs (circadian rhythm)\citep{Gaudreau1994}. %\citep{Kang2013,Gaudreau1994}. 
 Photosynthesis is best when incident light, CO$_2$ concentration, and temperature are at levels optimal for the plant growth. The model equation \eqref{eq:dB/dt} only describes the plant growth due to photosynthesis while the circadian rhythm is introduced through the reference trajectories.
	
% Photosynthesis, and subsequently plant growth, is higher with better utilization of incident light, when the temperatures and CO$_2$ concentrations are at levels optimal for the plant. 
 %This, however, is also influenced by light and temperature dynamics due to the circadian rhythm of nature \citep{Kang2013,Gaudreau1994}, which is not inherently modeled.
 
 An approximated reference for the daily light input trajectory $u_{\mathrm{I}j}^{\text{ref}}$, $j=1,\dotsc,4$ is specified using a cosine function. In particular
 \begin{align*}
 u_{\mathrm{I}j}^{\text{ref}}(t) = 50 - 50 \cos(2 \pi f_{\mathrm{Hz}}t), \quad j = 1,\dotsc,4,
 \end{align*}
 where $f_{\mathrm{Hz}}= 1/(2\cdot 60 \cdot 24) = 1/\mathrm{Day}$.  
 Furthermore, a cosine trajectory is adopted for $T^{\text{ref}} \text{ and } C^{\text{ref}}$ such that the times of peak values in light intensity, temperature and CO$_2$ concentration coincide. 
The near optimal reference is suggested as
\begin{align*}
T^{\text{ref}}(t) &= 20 - 3 \cos(2 \pi f_{\mathrm{Hz}}t),\\
C^{\text{ref}}(t) &= \num{9.05e-4} - \num{1.8e-04} \cos(2 \pi f_{\mathrm{Hz}}t).
\end{align*}
 
Regarding the input reference values,
\begin{align*}
	u_{\mathrm{T}}^{\text{ref}}(t) &= u_{\mathrm{V}}^{\text{ref}}(t) = u_{\mathrm{H}}^{\text{ref}}(t) = u_{\mathrm{W}1}^{\text{ref}}(t) = u_{\mathrm{W}2}^{\text{ref}}(t)= u_{\mathrm{W}3}^{\text{ref}}(t) 
	\equiv 0
	%C^{\text{ref}}(t) &\equiv \num{9.05e-4} - \num{1.8e-04} \cos(2 \pi f), \\
	%H^{\text{ref}}(t) &\equiv f_H(60 \% \text{ relative humidity}, T^\mathrm{ref}(t))\;,\\
	%W_\mathrm{sto}^{\text{ref}}(t) &\equiv 0.1, \, W_\mathrm{med}^{\text{ref}}(t) \equiv 0.5, \, W_\mathrm{ovf}^{\text{ref}}(t) \equiv 0.5, \; B^{\text{ref}}(t) \equiv 0   
\end{align*}
is used, as to capture the value of minimum energy expenses.
For this study, the disturbance trajectory was generated using records of past weather data of Chemnitz, Germany. 
%The disturbance is deliberately chosen to contain significantly higher temperature values which may lead to reaching the actuator limitations. 

%Since no terminal constraint on the temperature at the end of the prediction horizon is included in the optimization in \eqref{eq:MPC}, the penalty of the error between predicted state and input and their respective reference is set sufficiently high. 
For tracking, respective state and input weights are set sufficiently high.
Specifically, the stage cost function
\begin{equation}
\begin{split}
	r(&\mathbf{x}(t_k), \mathbf{u}(t_k)) = 5000 \, ( \, T(t_k) - T^{\text{ref}} (t_k+i) \, )^2 \,\\
	& + \num{1.11e12} \, ( \, C(t_k) - C^{\text{ref}} (t_k+i) \, )^2 \,\\
	& + \left(\mathbf{u}_{i}(t_k)-\mathbf{u}^{\text{ref}}(t_k+i) \right)^\top R \left( \mathbf{u}_{i}(t_k)-\mathbf{u}^{\text{ref}}(t_k+i) \right),\\
\text{with } &R = \text{diag}(0.1,1,0.25,0.5,0.5,0.5,100,\dotsc,100).
\end{split}
\end{equation}
%is employed, using 
%\begin{equation*}
%R = \text{diag}(0.1,1,0.25,0.5,0.5,0.5,100,\dotsc,100).
%\end{equation*}
The values of state penalties are chosen to compensate the different scales of various states. In turn the actuation costs for all actuators except LEDs are equivalent to the current consumed in amperes. For example, when $u_\mathrm{T}=50$, 5\si{\ampere} of current is consumed by the TEC.

%Since state trajectories of temperature and C0$_2$ concentration have significant influence on the plant growth, respective error weights in the stage cost are specified as non-zero. 

%% file: results_discussion.tex
\section{Results and Discussion}\label{sec:result}
\begin{figure*}[!h]
	%\begin{center}
	%\includegraphics[width=13cm]{gfx/Plant_process_growth_chamber_ext}    % The printed column width is 8.4 cm.
	%		\resizebox{\linewidth}{!}{\input{./gfx/Plant_process_growth_chamber_ext.pdf_tex}}
	%\def\svgwidth{\linewidth}
	%\fontsize{7.5}{8}\selectfont
	\centering
	\includegraphics[trim=20 273 10 273,clip,scale=0.9]{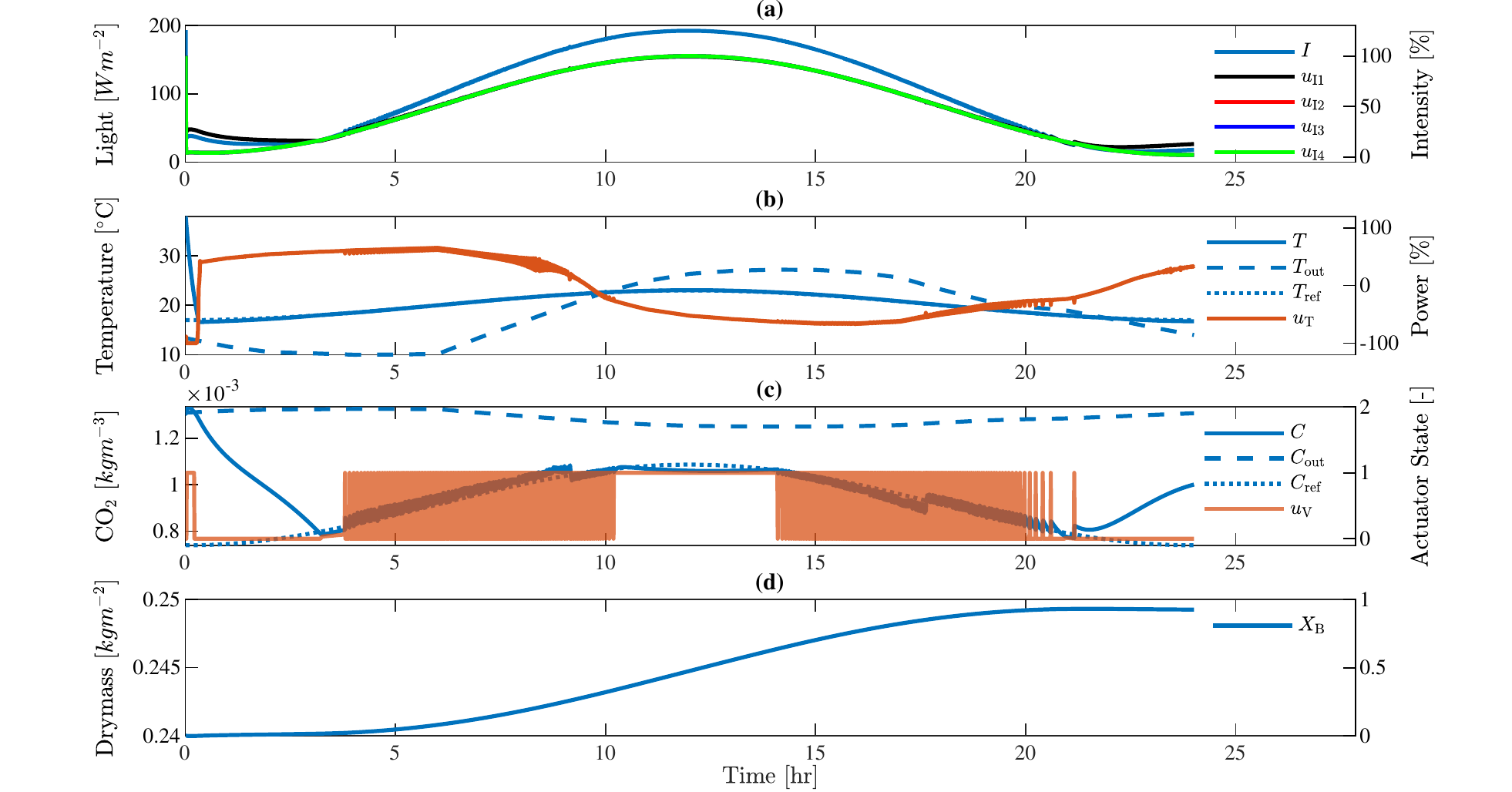}
	\caption{Simulation results: The model was simulated with the implemented MPC for a duration of 24 hrs. The state trajectories, references, disturbances and the corresponding inputs are visualized. Chattering of certain actuators \eg the ventilator, are non-crucial for the production unit used in this work (switching frequency $< 0.0333$ Hz).}
	\label{fig:ctrl_perf}
	%\end{center}
\end{figure*}

The results of the simulation, which is carried out for 24 hrs, can be seen in Fig. \ref{fig:ctrl_perf}. 
The simulation starts with the initial state vector
\begin{equation*}
\mathbf{x}(t_0) = 
\begin{bmatrix} 38 & 0.0013 & 0.0058 & 0.0 & 0.0 & 0.0 & 0.240
\end{bmatrix}^\top,
\end{equation*}
with $t_0$ being midnight. The starting weight of the plant, corresponding to lettuce size ready for harvest, and a high initial temperature are considered for maximum operation load on the actuators. 

At first, one can observe constraint satisfaction for all states and inputs according to the specification in the control optimization.
The light intensity follows the given reference sinusoid pattern reaching the peak amplitude at noon \ie at 12 hr, for all LED channels (see Fig.~\ref{fig:ctrl_perf}(a)).

Simultaneously, temperature inside the chamber reaches its reference in 0.3 hrs and is able to maintain its reference trajectory for the entire time (see Fig.~\ref{fig:ctrl_perf}(b)).
Jitters in the control signal for the TEC can be noticed at the beginning, around 5-10 hr, and 15-20 hr.
These jitters are controller's response to temperature fluctuation due to the activation of the ventilator (see Fig.~\ref{fig:ctrl_perf}(c)). % which brings in the outside air with temperature corresponding to $T_\mathrm{out}$. 
It was also observed from simulations with higher $T_\mathrm{out}$ temperatures such that the difference $T_\mathrm{out}-T^\mathrm{ref} > 10~^\circ$C, temperature tracking was not achievable.
%This can be explained by the heat transfer capacity (cooling) of the actuator, which is limited to cooling the chamber not more than $10^\circ$C below the ambient temperature (as mentioned previously in Section \ref{sec:modeling}).
This can be explained by the limitation in heat transfer capacity (cooling) of the actuator(as mentioned previously in Section \ref{sec:modeling}).  

CO$_2$ concentration in the chamber increases at the beginning and at the end of the simulation due to respiration. Since it is not possible to reduce this concentration through ventilation at the mentioned times, the controller increases the light intensity activating photosynthesis and thus CO$_2$ consumption. At other times, the CO$_2$ concentration is tracked by frequent switching of the ventilator, providing CO$_2$ from the outside (see Fig.~\ref{fig:ctrl_perf}(c)). In the range 0.025-0.2 hr, ventilator is activated to accelerate cooling and thus reach the reference temperature value. It can be noticed that once the light intensity elevates, the CO$_2$ consumption due to the plant photosynthesis is at maximum requiring constant CO$_2$ flow.
The slew rate used for the ventilator is acceptable for the production unit used in this work. However, for systems with limitation in switching rate, the control problem needs to be modified such that high frequency switching is penalized.

%by the ventilator, exchanging the gas with the outside source at higher concentration. The ventilator is mostly kept ON to enable the gas exchange during the entire simulation period. It can be noticed that once the light intensity elevates, the CO$_2$ consumption due to the plant photosynthesis reduces the actual concentration in the chamber approaching that of the reference value (see Fig.~\ref{fig:ctrl_perf}(c)). %Another hardware limitation can be observed from this simulation. The hardware setup enables the gas concentration regulation by exchanging with the outside source which works only when the outside concentration is favorable.

%Although, humidity is not tracked it follows the pattern of the temperature, rising gradually with the temperature reaching its peak (see Fig.~\ref{fig:ctrl_perf}(d)). 
%Humidity inside the chamber varies, following the pattern similar to temperature, as a result of transpiration from the surface of the leaves. As the humidity approaches the reference, between 1 hr and 8 hr, humidifier and ventilator is activated repeatedly to bring the humidity close to the reference trajectory(see Fig.~\ref{fig:ctrl_perf}(d)). After 8 hr, ventilator is constantly ON while the humidifier is turned OFF, constantly exchanging the humid air and increasing the humidity inside the chamber. At 17 hrs, when the outside humidity falls below the reference trajectory, humidifier is activated again and stays ON for the duration the humidity difference between $H_\mathrm{out}$ and $H^\mathrm{ref}$ is small.
Water levels in the tanks are maintained such that the state constrains (tank capacities) posed on these water levels are not violated. % (see Fig.~\ref{fig:ctrl_perf}(e)).
The biomass growth shown in Fig.~\ref{fig:ctrl_perf}(d) appears to be highest between 7-17 hr when the conditions for growth (temperature and CO$_2$ concentration) are optimal. Beyond these time points, biomass production rate is minimal. 
%Restoration of favourable CO$_2$ concentration is possible by activating the ventilator. This, however, also affects the inside temperature which subsequently adjusts to the outside value.

These simulations depicts certain capability of the tracking control, that could satisfy all state and input constraints according to the specification within the control optimization \eqref{eq:MPC}, while also pointing out particular challenges.